\documentclass[sigconf]{acmart}
\AtBeginDocument{%
  }

\copyrightyear{2026}
\acmYear{2026}
\setcopyright{cc}
\setcctype{by}
\acmConference[SIGIR '26] {Proceedings of the 49th International ACM SIGIR Conference on Research and Development in Information Retrieval}{July 20--24, 2026}{Melbourne, VIC, Australia}
\acmBooktitle{Proceedings of the 49th International ACM SIGIR Conference on Research and Development in Information Retrieval (SIGIR '26), July 20--24, 2026, Melbourne, VIC, Australia}
\acmISBN{979-8-4007-2599-9/2026/07}
\acmDOI{10.1145/3805712.3809752}
\usepackage{amsmath}
\usepackage{booktabs}
\usepackage{pbox}
\usepackage{multirow}
\usepackage{graphicx}
\usepackage{tcolorbox}
\usepackage{pbox}
\usepackage{multirow}
\usepackage{makecell} 
\usepackage{arydshln}
\usepackage{xcolor}

\tcbset{
  myprompt/.style={
    colback=gray!5,       
    colframe=gray!50,      
    fonttitle=\bfseries,   
    coltitle=black,        
    boxrule=0.5mm,         
    width=.45\textwidth,      
    top=5pt, bottom=1pt,   
  }
}
\newcommand{\B}[1]{\textbf{#1}}

\begin{document}

\title{Search for Coverage: Learning Coverage-Aware Retrieval with Augmented Sub-Question Answerability}

\author{Jia-Huei Ju}
\affiliation{%
  \institution{University of Amsterdam}
  \country{Amsterdam, Netherlands}
}
\email{j.ju@uva.nl}

\author{Eugene Yang}
\affiliation{%
  \institution{Johns Hopkins University}
  \country{Baltimore, MD, United States}
}
\email{eugene.yang@jhu.edu}

\author{Trevor Adriaanse}
\affiliation{%
  \institution{Johns Hopkins University}
  \country{Baltimore, MD, United States}
}
\email{tadriaa1@jhu.edu}

\author{Suzan Verberne}
\affiliation{%
  \institution{Leiden University}
  \country{Leiden, Netherlands}
}
\email{s.verberne@liacs.leidenuniv.nl}

\author{Andrew Yates}
\affiliation{%
  \institution{Johns Hopkins University}
  \country{Baltimore, MD, United States}
}
\email{andrew.yates@jhu.edu}


\renewcommand{\shortauthors}{Jia-Huei Ju, Eugene Yang, Trevor Adriaanse, Suzan Verberne, \& Andrew Yates}

\begin{abstract}
Long-form Retrieval-Augmented Generation (RAG) brings the challenge of coverage-based ranking, because ranking methods must ensure the inclusion of comprehensive relevant nuggets (i.e., facts), which can thereby be synthesized into a comprehensive output. In this work, we propose CoveR,\footnote{Our code is available at https://github.com/DylanJoo/CoveR} a dense retrieval method optimized for coverage-aware retrieval scenarios.  CoveR is a bi-encoder trained with the coverage-based contrastive and distillation objectives, which enables CoveR to capture diverse aspects of information needs. 
To train CoveR, we create the SCOPE dataset,\footnote{Our training data is available at https://huggingface.co/datasets/DylanJHJ/scope} which comprises 90K training pairs from Researchy Questions with synthetic coverage signals augmented from sub-question answerability judgments generated by LLMs. Our empirical experiments show that CoveR enhances nugget coverage by 10\% over strong dense retrieval baselines without sacrificing its relevance-based retrieval capability. 
Further ablation studies validate the importance of our proposed learning method, showing that CoveR achieves a superior trade-off between relevance- and coverage-based ranking, which is essential for long-form RAG.

\end{abstract}




\maketitle

\section{Introduction}
\begin{figure}
    \centering
    \includegraphics[width=.9\columnwidth]{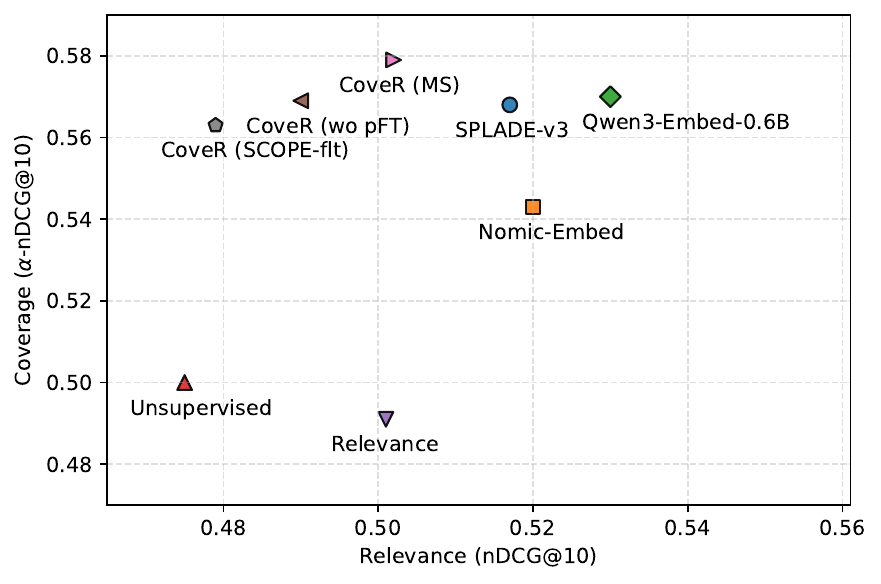}
    \caption{
    Effectiveness for relevance ($x$-axis) vs. information coverage ($y$-axis). 
    Relevance is reported with the average nDCG@10 across 13 BEIR datasets. 
    Information coverage is reported with the average $\alpha$-nDCG@10 across 3 nugget-based retrieval evaluation datasets.
    }
    \label{fig:two-metrics-cov}
\end{figure}

As LLMs have evolved to support longer contexts, a new search paradigm has emerged. Beyond presenting search results via a simple ranking list, recent search engines have integrated LLM generation for synthesizing search results into a structured report with citations~\cite{Mayfield2024-vw} (e.g., Google's Search and its AI Mode).
This paradigm is also known as one of the long-form Retrieval-Augmented Generation (RAG) tasks~\cite{Stelmakh2022-kt, Gao2023-oh, Tan2024-ff}, where the input query may consist of multiple sub-information needs, and the final output is expected to be comprehensive.
To this end, the primary goal of retrieval shifts from finding the most relevant document to ensuring the comprehensiveness of nuggets (i.e., relevant facts), placing new demands on retrieval models to identify a set of documents that can comprehensively cover diverse aspects of the user's information need.

This shift has motivated a reconsideration of how retrieval should be evaluated. Recent studies have begun to assess retrieval of long-form RAG through the lens of information coverage~\cite{Ju2025-oz, Samarinas2025-dg} at a more fine-grained nugget level~\cite{Voorhees2003-ul, Min2021-xr}.
Notably, such evaluation highlights the overlooked drawbacks of redundancy and lack of diversity in RAG~\cite{Chen2024-sp}, as the increase of document-level relevance translates only limitedly into gains in nugget coverage~\cite{Ju2025-oz}, implying top-ranked documents may look different while containing similar nuggets. This again emphasizes the critical demand of coverage-aware retrieval.

However, standard neural retrievers are predominantly trained with relevance-based supervision like MSMARCO passage ranking~\cite{bajaj2016ms}, which encourages queries and their relevant documents to cluster in a narrow region in the embedding space. 
While it is effective for relevance-based ranking, this embedding geometry is sub-optimal for the retrieval scenario of long-form RAG like report generation tasks~\cite{Mayfield2024-vw}.
To be more specific: 
(a) on the query side, a single query often comprises multiple diverse and open-ended information needs (see Table~\ref{tab:scope-example}), making it challenging to ensure retrieved documents cover a comprehensive set of nuggets;
(b) On the document side, the narrow region of the retrieval scope leads to favoring documents that are highly relevant while containing similar facts.
As a result, addressing this representation challenge of relevance optimization and coverage-awareness is essential for advancing retrieval in the context of future search systems.

In addition to the representation challenge, another obstacle for coverage-aware retrieval is the availability of a suitable training dataset. 
Commonly used datasets for relevance ranking, such as MSMARCO~\cite{bajaj2016ms} or NQ~\cite{kwiatkowski-etal-2019-natural}, provide supervision signals where each training query has a short-form answer attached to it.
This implies a narrow definition of information need in their training query, typically resembling the retrieval for short-form question answering~\cite{Lewis2020-ky, Karpukhin2020-wc}, which requires one or only a few information nuggets.
In contrast, the query collection from Researchy Questions~\cite{Rosset2024-hq} focuses on broader queries that demand a deeper understanding of information needs; this aligns well with the notions of coverage-aware retrieval for long-form RAG.
Moreover, the dataset also includes decomposed sub-questions that explicitly reflect diverse views of the query. 
These sub-questions can then be naturally augmented as coverage signals by employing nugget-level retrieval evaluation framework~\cite{Ju2025-oz}.

In this work, we introduce CoveR, a \textbf{Cove}rage-aware \textbf{R}etriever with tailored coverage-based training methods: \emph{Coverage contrastive} and \emph{Coverage self-distillation}.
We instantiate the coverage contrastive signals by sampling positive and negative documents according to their coverage scores. This objective helps tweak the initial relevance-aware embedding space towards considering multiple views in the query, enabling the predicted similarity score to reflect diverse nuance between documents.
On top of that, we use the additional synthetic sub-questions to assist the training of query encoding with self-knowledge distillation~\cite{Chen2024-hh}.
For each query, we aggregate similarity scores from multiple sub-questions into the augmented coverage score, which is considered as a teacher score for the predicted score using the original query.

To support training CoveR, we create SCOPE, a training dataset with augmented coverage signals.
We curate SCOPE using training queries from Researchy Questions~\cite{Rosset2024-hq}, leveraging their inherent structure of a query and its associated sub-questions.\footnote{\url{https://huggingface.co/datasets/corbyrosset/researchy_questions}}
However, the original data lacks relevance judgments linked to each decomposed sub-questions, which limits the usability of the collection as a new training resource.
We mitigate this shortcoming by labeling relevant documents with different grades through sub-question answerability~\cite{Sander2021-lg, Farzi2024-yn} using Llama-3 70B model~\cite{Grattafiori2024-zw}.

To investigate the connection between our coverage-aware retrieval model and nugget coverage metrics like $\alpha$-nDCG, we conduct experiments on the NeuCLIR report generation benchmark dataset~\cite{Mayfield2024-vw} and CRUX evaluation datasets~\cite{Ju2025-oz}. We find that CoveR with coverage-based training on SCOPE can substantially outperform comparable baselines with standard relevance training, including the same backbone model trained on MSMARCO.
Our empirical evaluation on BEIR also showcases that pre-finetuning on MSMARCO is important for balancing coverage and relevance ranking effectiveness, as depicted in Figure~\ref{fig:two-metrics-cov}.

Our contributions are:
\begin{itemize}
    \item We propose the CoveR model, a bi-encoder trained to improve nugget coverage using a contrastive coverage-aware loss or a coverage self-distillation loss.
    \item We create the SCOPE dataset for training coverage-aware retrieval models by augmenting Researchy Questions with sub-question answerability judgments and generating synthetic queries with multiple aspects.
    \item We conduct extensive experiments on collections with nugget judgments that demonstrate that CoveR improves nugget coverage over comparable baselines without harming relevance metrics. Furthermore, we demonstrate on the BEIR benchmark that CoveR continues to perform well on standard benchmarks that consider document relevance.
\end{itemize}

\begin{figure*}
    \centering
    \includegraphics[width=1.9\columnwidth]{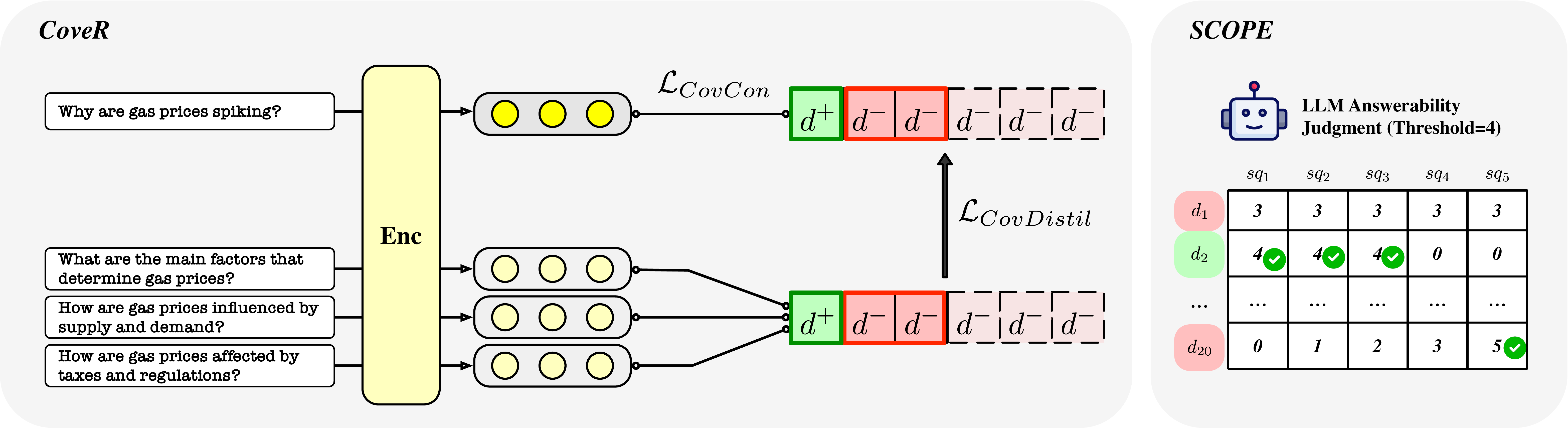}
    \caption{The two proposed coverage-based training methods: coverage contrastive (CovCon) and coverage self-distillation (CovDistil). The sampled positive and negative documents are selected based on coverage scores.}
    \label{fig:CoveR}
\end{figure*}

\section{Related work}
With the advances in representation contrastive learning~\cite{Karpukhin2020-wc, Lee2019-nt}, neural dense retrieval models have achieved great success in relevance ranking tasks, such as MSMARCO passage ranking~\cite{bajaj2016ms} or TREC DL~\cite{craswell2025overviewtrec2021deep}.
Particularly, dense retrieval learns to represent the query and documents with embeddings contrastively, and thereby estimates the relevance by geometric distance.
Over this decade, many researchers have refined retrieval models through finer-grained negative sampling strategies~\cite{xiong2020approximatenearestneighbornegative}, knowledge distillation from cross-encoders~\cite{hofstätter2021improvingefficientneuralranking}, or scaling embedding models, training data, and training time~\cite{Chen2024-hh, qu2021rocketqa, lin2023traindragondiverseaugmentation}. 

However, dense retrieval may be vulnerable when adapting to shifted domains or specialized retrieval tasks~\cite{Thakur2021-cc}.
This becomes increasingly pronounced due to the recent emergence of long-form RAG applications~\cite{Gao2023-oh, Stelmakh2022-kt}, which has redefined the demands on retrieval.
Shifting from the traditional relevance ranking, the new search engines have started to integrate LLMs in the system~\cite{Mayfield2024-vw}.
In the new search paradigm, retrieval models require to maximize the ``information coverage'', so that the downstream generator can resolve the more complicated information needs~\cite{Dawn2025-sg}.
These changes allow users to query more complex information need where the definition of relevance becomes compounded~\cite{yang2024cragcomprehensiverag}, shifting the retrieval goal from finding the most relevant documents to achieving comprehensive coverage across multiple documents~\cite{Ju2025-oz, Samarinas2025-dg}.

To support this development, many recent studies have revisited nugget-level evaluation~\cite{Voorhees2003-ul, Grusky2018-ss, Fabbri2019-hc} that go beyond traditional evaluation protocol via document-level relevance judgments.
A ``nugget'' is defined as a fact for which the assessor could make a binary decision as to whether a document contained the fact, which is well-aligned to the goal of improving coverage of retrieval for long-form RAG~\cite{Mayfield2024-vw}.
To facilitate a more informed reflection on potential retrieval model designs for coverage,~\citet{Ju2025-oz} proposed an evaluation framework to measure coverage-based metrics: $\alpha$-nDCG~\cite{Clarke2008-cn} and coverage. 
We hypothesize that the coverage-aware retrieval capability is a major obstacle for long-form RAG scenario, as the standard retrievers are optimized for ranking individual documents~\cite{Wechsler2000-ai} rather than ensuring comprehensive coverage of all sub-information needs.

Although there are limited existing retrieval approaches that are particularly tackled for coverage, some prior works on retrieval diversification have explored similar thoughts. 
For example, techniques such as Maximal Marginal Relevance (MMR)~\cite{Carbonell1998-gf}, query reformulation~\cite{Li2024-aa}, and multi-query retrieval or reranking~\cite{Zhong2025-wy, ju2026lancer} have been used to reduce redundancy and increase result diversity~\cite{macavaney2021intent5searchresultdiversification}, thereby benefits RAG performance~\cite{Wang2025-ym}.
Some recent research has started exploring new ranking methods that can retrieve a more comprehensive set of relevant information~\cite{Lee2025-xm}, or generate multiple query embeddings for diverse views~\cite{chen2025singleembeddingscapturingdiverse}, which could collectively support downstream synthesis of generators.

\section{Coverage-Aware Retrieval}
In this section, we introduce our proposed coverage-aware retriever, CoveR. 
First, we describe the bi-encoder architecture. 
Second, we introduce two coverage-based training objectives, aiming at reflecting the volume of coverage on the similarity scores.
Last, we describe SCOPE, a special training data with augmented coverage signals, which is used for training CoveR.

\subsection{Bi-encoder Architecture}\label{sec:bi-encoder}
As illustrated in Figure~\ref{fig:CoveR}, CoveR adopts the standard bi-encoder architecture where the query and document are represented as a dense vector each via an encoder ${\rm Enc}_\theta$.
Queries and documents are independently encoded and are concatenated with the prefix as follows:
\begin{align}
    E_q &= {\rm Enc}_\theta(\texttt{``search\_query: }\{q\}\texttt{''}); \notag \\
    E_d &= {\rm Enc}_\theta(\texttt{``search\_document: }\{d\}\texttt{''}), \label{eq:prefix}
\end{align}
where the $E$ represents the contextualized token embeddings. 
With them, we can then calculate the score of each query-document pair with similarity function such as cosine:
\begin{align}
    s(q, d) = \dfrac{  {\rm Mean}(E_q) \cdot {\rm Mean}(E_d) }{ 
    \|{\rm Mean}(E_q)\| \|{\rm Mean}(E_d)\| }, \notag
\end{align}
where each the query and document is first represented in a mean-pooled vector and normalized for the integration of nearest neighbor search infrastructure (i.e., FAISS).

\subsection{Learning with Sub-Questions}
Modern bi-encoders are typically trained with either a contrastive learning objective or a distillation objective.
Contrastive learning has the benefit of requiring only relevance labels, whereas distillation objectives can often result in higher model effectiveness, although they require scores from more expensive teacher models.

Inspired by these two paradigms, we introduce a coverage-based contrastive learning objective dubbed CovCon and a coverage-based self-distillation objective dubbed CovDistill. 
As depicted in Figure~\ref{fig:CoveR}, both objectives use the decomposed sub-questions to train CoveR to rank documents based on \emph{how many relevant nuggets they contain} rather than \emph{how relevant the documents are overall}.

\subsubsection{CovCon: Coverage Contrastive Learning}\label{sec:covcon}
Relevance-based neural retrievers are often trained using a contrastive objective. Geometrically, such objectives use the query $q$ as an anchor embedding to pull positive samples $d^+$ closer to it while pushing negative samples $d' \in D^-$ away, including the in-batch negatives~\cite{Yih2011-qd, henderson2017efficientn}. 
This can be implemented with a softmax-normalized cross entropy with a temperature $t$ like:
\begin{align}
    \mathcal{L}_{\rm CovCon} = 
    - \log \frac{\exp\big(s(q, d^+) / t\big)}
    {\sum_{d' \in \{d^+, D^-\}} \exp\big( s(q, d') / t \big)}.\label{eq:cc-crossentropy}
\end{align}
To achieve our goal of coverage-based ranking, we redefine the definition of similarity by selecting positive and negative samples based on coverage score.
Specifically, for a query $q$, we sample a positive $d^+$ from the group of relevant documents that has high coverage scores, denoted as $D_{HC}$ and multiple negatives from the group of documents that has low coverage scores:
\begin{align}
    D_{HC} \leftarrow &  \{ d \in \mathcal{D} \mid Cov(q, d) \in [\alpha, \alpha') \}; \notag \\
    D_{LC} \leftarrow &  \{ d \in \mathcal{D} \mid Cov(q, d) \in [\beta , \beta')\}, \label{eq:cov-sampling}
\end{align}
where $\mathcal{D}$ is a set of documents retrieved using BM25 with $q$ as query (see Section~\ref{sec:scope} for details). 
$Cov(q, d)$ indicates the coverage scores of the document $d$ given the query $q$. We calculate the coverage score defined in prior work with sub-question answerability~\cite{Ju2025-oz} (i.e., how many query-associated sub-questions are answered with each documents).
Parameters $\alpha$ and $\beta$ control the range of coverage scores. The impact of the different sampling range is reported in our ablation analysis in Section~\ref{sec:ana-cov-sample}.

\subsubsection{CovDistil: Coverage Self-Distillation}
To reinforce the coverage-awareness in query encoding, we introduce an efficient self-distillation process.
Pre-computing coverage scores from every combination of sub-questions and documents would be computationally expensive; instead, CovDistil repurposes the estimated similarity scores for sub-questions as a naturally available distillation target. 
During training, as the document embeddings are encoded, we only need to encode the sub-question embeddings for computing the teacher scores.

As illustrated in Figure~\ref{fig:CoveR}, we construct a teacher score distribution by aggregating similarity scores across all sub-questions $sq_j \in SQ$. 
Each sub-questions $sq$ is similarly encoded as query $q$, then interacts with all documents $\mathbf{d}$ in the mini-batch (the positives and the negatives from all the other queries). The resulting scores are average similarity scores across sub-questions to form the teacher score distribution:
\begin{align}
    P_{sq}(\mathbf{d}|q) &= \dfrac{\exp \Big( \mu_{sq\in SQ}\big( s(sq, d^+) \big)  / t\Big)}
    {
        \sum_{d' \in \{d^+, D^-\}} \exp \Big( 
            \mu_{sq\in SQ}\big( s(sq, d') \big) / t \Big)}, \label{eq:cd-teacher}
\end{align}
where $\mu_{sq\in SQ}(sq, d)$ denotes the mean values over all sub-question similarity for the document $d$. 
Finally, we employ Kullback-Leibler (KL) divergence to align the student score, calculated using the original query $q$, with the teacher score distribution $P_{sq}(\mathbf{d}|q)$.
Both score distributions are calculated with the same encoder and documents; this not only preserves the encoder's initial capability but adds the coverage-awareness on top of it:
\begin{align}
    \mathcal{L}_{\rm CovDistil} = 
    \lambda_{CD} \cdot \mathbf{KLDiv}\big( P(\mathbf{d}|q) || P_{sq}(\mathbf{d}|q) \big), 
\end{align}
where $P(\mathbf{d}|q)$ indicates the estimated coverage score distribution using query $q$ (i.e., student scores), which is from the Eq.~\eqref{eq:cc-crossentropy}.
For each query, we combine the CovCon as well as the CovDistil with a weight $\lambda_{CD}$.
More analysis is reported in Section~\ref{sec:ablation}.

This objective serves as a regularizer to stabilize the transition from standard relevance-based bi-encoders to coverage-based. It aims to provide a smoother gradient update that prevents \emph{relevance collapse}--a phenomenon where the encoder loses its initial relevance estimation capabilities due to the shift in how positive and negative samples are defined in Eq.~\eqref{eq:cc-crossentropy}.

\subsubsection{Relevance Pre-finetuning}\label{sec:rel-pft}
Prior to coverage-based training, we found in pilot experiments that it is beneficial to first warm up bi-encoders with relevance-based training datasets.
The motivation is to ensure that the prior embedding space secures satisfactory representation capability, thereby achieving a more ideal self-distillation process.
Specifically, we pre-finetune CoveR with the standard relevance-based contrastive learning identical to the Eq.~\eqref{eq:cc-crossentropy} but different (external) training pairs.
We use the MSMARCO passage ranking dataset for relevance pre-finetuning.
The negative samples are mined from the top-100 BM25 and dense retrieval models.

\begin{table}[t]
    \centering
    \caption{An example of Researchy Questions and SCOPE dataset. 
    $q$ is the original user query while $q'$ is the synthesized query made for covering the aspects in the set of decomposed sub-questions $SQ$, which are provided in the original dataset.}
    \label{tab:scope-example}
    \resizebox{\columnwidth}{!}{
    \begin{tabular}{ll}
        \toprule
        RyQ: $q$ & Why are gas prices spiking? \\
        \midrule
        RyQ: $sq_j \in SQ$ &  
        \pbox{7cm}{
        - What are the main factors that determine gas prices? \\
        - How are gas prices influenced by supply and demand? \\
        - How are gas prices affected by taxes and regulations \\
        ...}  \vspace{0.1em} \\
        \midrule
        SCOPE: $q'$ & 
        \pbox{7cm}{
        Produce a report on the factors that determine gas prices in the United States. The report should provide an in-depth analysis of the current trends, ... as well as the underlying factors that influence gas prices ... } \vspace{0.1em} \\
        \bottomrule
    \end{tabular}
    }
\end{table}

\subsection{The SCOPE Coverage Training Dataset}\label{sec:scope}
There is no dataset with large-scale nugget coverage labels for coverage-based training. We therefore build SCOPE, a training dataset with coverage signals, which consists of 90K synthetic coverage training pairs. 

\subsubsection{Query with Multiple Aspects.}
Recently,~\citet{Rosset2024-hq} released \emph{Researchy Questions}, which consists of 90K training queries taken from the Bing query log, along with the web documents that the user clicked.
Research-type queries usually require searching with multiple sub-queries in a single session to satisfy multiple aspects of the information need.
In the original dataset, the authors provide the decomposition of each query into multiple sub-questions generated by GPT4. 
Each decomposed sub-question can be naturally regarded as the proxy of an information nugget and fit our goal of optimizing coverage. 
In our preliminary experiments, however, we observe that the original query has insufficient semantic connections to multiple sub-questions, which results in a misalignment between the query and sub-questions.
To mitigate this, we re-generate a more aligned request-like query using an in-context prompt, as shown in Table~\ref{tab:scope-example}.

\subsubsection{Candidate Relevant Documents}
Relevance labels in Researchy Questions are based on documents users clicked after issuing the original query.
Such sparse labels are insufficient for developing coverage-awareness, as they are only indirectly connected to a few sub-questions; the documents relevant to a specific sub-question are not necessarily labeled.
To be more usable as a new training resource for coverage-aware retriever, we collect additional pseudo-relevant documents.

First, we retrieve the top-100 candidate documents using BM25 from the Clueweb Category-B corpus~\cite{overwijk2022clueweb22}.\footnote{The corpus is the subset of the corpus used in Researchy Question~\cite{Rosset2024-hq}}
This retrieval serves as candidate document selection to make the following process computationally feasible.
Second,  
we use an instruction-tuned Qwen3 reranker\footnote{\texttt{Qwen/Qwen3-Reranker-0.6B}} to rerank the top-100 retrieved documents using the modified instruction shown in Figure~\ref{fig:rerank}.
This reranking aims at pushing the documents that are more relevant to multiple sub-questions to the top, making the limited labels more informative for training.\footnote{The \textit{join(sub-questions)} function produces a list of sub-questions}
Last, we select the top-20 documents as candidate relevant documents. The clicked documents supplied by the original datasets (439,151 in total) are also included as relevant documents to minimize the number of false positives in the corpus.

\begin{figure}[t]
\begin{tcolorbox}[title=Qwen3-Reranking Instruction, myprompt]
    \textbf{System prompt}: 
    Judge whether the Document meets the requirements based on the Query and the Instruct provided. Note that the answer can only be "yes" or "no". \\
    \textbf{Prompt}: 
    \textless Instruct\textgreater: Given the list of questions as query, retrieve relevant passages that answer the questions.
    \textless Query\textgreater: \{{\rm join}(sub-questions)\}
    \textless Document\textgreater: \{document\}"
\end{tcolorbox}
\caption{The instruction prompt used for generating distillation scores. The sub-questions are generated with GPT-4 from the original ResearchyQuestion. The candidate documents are retrieved from Clueweb category-B using BM25.}
\label{fig:rerank}
\end{figure}

\begin{figure}[t]
\begin{tcolorbox}[title=Document Answerability Judgment~\cite{dietz2024workbench}, myprompt]
Instruction: Determine whether the question can be answered based on the provided context? Rate the context with on a scale from 0 to 5 according to the guideline below. Do not write anything except the rating. \\\\
Guideline: \\
5: The context is highly relevant, complete, and accurate. \\
4: The context is mostly relevant and complete but may have minor gaps or inaccuracies.\\
3: The context is partially relevant and complete, with noticeable gaps or inaccuracies.\\
2: The context has limited relevance and completeness, with significant gaps or inaccuracies.\\
1: The context is minimally relevant or complete, with substantial shortcomings.\\
0: The context is not relevant or complete at all. \\\\
Question: \{$q$\} \\ Context: \{$c$\} \\ Rating:
\end{tcolorbox}
\caption{Rubric-based answerability judgment prompt. The output rating is converted into 0 to 5, and the output with incorrect formats is assigned to 0.}
\label{fig:judge}
\end{figure}

\subsubsection{Automatic LLM Judgments}
After identifying candidate relevant documents, we used the Llama3.3 70B model~\cite{Grattafiori2024-zw} to produce relevance judgments.
Each of the top-20 candidate relevant documents is automatically judged for each sub-questions from the original dataset.
Judgments are on a 0-5 scale using the prompt from~\citet{Sander2021-lg}, which are intended to assess the answerability~\cite{dietz2024workbench, farzi2024exam} of a question given a document (see next section).
The prompt is shown in Figure~\ref{fig:judge}.
As a result, we generate 24M judgments for the 1.2M sub-questions from Researchy Questions, as reported in Table~\ref{tab:scope-data}.
We observe that the judgments of the selected top-20 documents are not evenly distributed.
On average, each sub-question has 7.96 documents judged higher than 3 and 10.9 documents judged lower.

\begin{table}[t]
    \centering
    \caption{The statistics of LLM judgments on SCOPE dataset.}
    \label{tab:scope-data}
    \resizebox{.8\columnwidth}{!}{
    \begin{tabular}{lrrr}
        \toprule
        Grade & Count    & Proportion (\%) & \# Judgments / $sq$ \\
        \midrule
        5     & 1,907,722 & 7.85  & 1.48 \\
        4     & 5,444,774 & 22.39 & 4.22 \\
        3     & 2,909,134 & 11.96 & 2.26 \\
        2     & 7,920,642 & 32.57 & 6.14 \\
        1     & 2,651,382 & 10.90 & 2.06 \\
        0     & 3,482,653 & 14.32 & 2.70 \\
        Others& 13        & 0.00  & 0.00 \\
        \midrule
        Total & 24,316,320 & 100.00 & 18.86 \\
        \bottomrule
    \end{tabular}
    }
\end{table}

\begin{figure}
    \centering
    \includegraphics[width=.8\linewidth]{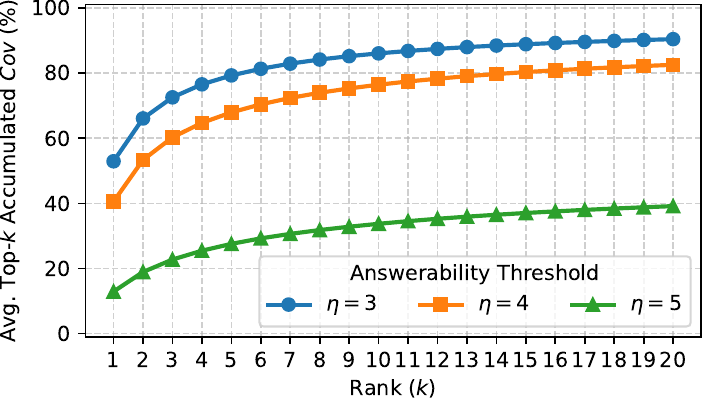}
    \caption{The accumulated Coverage scores of the top-$k$ selected relevant documents for different answerability thresholds.}
    \label{fig:topk-acc-cov}
\end{figure}

\subsubsection{Coverage-based Sampling}
Finally, we convert document judgments into binary answerability and then synthesize coverage scores. 
These scores serve as the criteria for sampling positives and negatives for coverage-based training methods (See Section~\ref{sec:covcon}).
Specifically, we set the threshold $\eta$ as 4 and then calculate the coverage score as:
$$
   \# \{sq_j \in SQ \mid J{(d, sq)} \geq \eta \} / |SQ|,
$$
where $SQ$ indicates a set of all decomposed sub-questions associated to a query. 
$J(\cdot)$ is the aforementioned LLM judgment for a sub-question and a document.
The score implies the proportion of sub-questions are answered.

Figure~\ref{fig:topk-acc-cov} depicts the coverage of answered sub-questions using top-1 to top-20 candidate documents, ranked from the previous stages.
Each curve represents the accumulated coverage under different answerability threshold $\tau$. We find that a threshold of 5 results in an overly strict answerability criterion, achieving less than 50\% coverage even with top-20 documents. 
Thresholds of 3 and 4 are both reasonable choices; however, a threshold of 4 can more cleanly separate documents into high and low coverage groups, which is assumably to be more desirable for our training process.

During training, we can thereby sample positive and negative documents from two groups $D_{HC}$ and $D_{LC}$ as described in Section~\ref{sec:covcon}.
We empirically test the different sampling strategies over varying $\alpha$ and $\beta$ in Eq.~\eqref{eq:cov-sampling}, and set $\alpha_1, \alpha_2=50\%, 75\%$ and $\beta_1, \beta_2 = -\infty, 0$.
For some cases that have fewer or no judged low-coverage documents, we supplement the negative group with the documents ranked below 50, resulting in 16 negatives for each query.
The resulting SCOPE training dataset is then constructed, with the summary statistics reported in Table~\ref{tab:datasets}.


\begin{table}[t]
    \centering
    \caption{Summary statistics for the SCOPE training dataset and for the nugget-based evaluation datasets used.
    The $\dagger$ indicates the nugget in NeuCLIR is built at the answer-level compared to the other, which are at question levels.}
    \label{tab:datasets}
    \resizebox{.97\columnwidth}{!}{
    \begin{tabular}{lrrrr}
        \toprule
        & \textbf{Train} & \multicolumn{3}{c}{\textbf{Evaluation}} \\
        \cmidrule(lr){2-2} \cmidrule(lr){3-5}
        & SCOPE
        & \makecell{NeuCLIR24\\ReportGen}
        & \makecell{CRUX-MDS\\DUC04} 
        & \makecell{CRUX-MDS\\Multi-News} \\
        \midrule
        \# Queries       & 81K & 19  & 50   & 100 \\
        \# Sub-Questions & 12M & 7K & 750  & 1K \\
        \# Documents     & 3M  & 10M & 565K & 565K \\
        \midrule
        \multicolumn{4}{l}{\bf Average Number Per Query} \\
        Nuggets         & 14.3   & 21.8$^\dagger$ & 15  & 10 \\
        \# Positive     & 5.1    & 89.8  & 31.9 & 8.1 \\
        \# Negatives    & 18.4   & -  & - & - \\
        \bottomrule
    \end{tabular}
    }
\end{table}

\section{Experimental Setup}
In this section, we first describe our implementations of the training for CoveR.
We then elaborate on the evaluation protocols for nugget-based and relevance-based retrieval benchmarks.
We also compare our proposed methods more broadly with other retrieval methods.

\subsection{Training}
In our experiments, we use the following training datasets for finetuning our internal model variants:
\begin{itemize}
    \item \textbf{MSMARCO Passage Ranking (MSMARCO)} consists of a large-scale collection of 8.8M passages and 491K training queries released by~\citet{bajaj2016ms}.
    Each query has one ``golden'' positive passage. We use the augmented dataset\footnote{https://huggingface.co/datasets/Tevatron/msmarco-passage-new} released by ~\citet{ma2025tevatron}, which each query has a set of hard negatives mined from a blend of BM25 and CoCondenser.
    \item \textbf{SCOPE} is the dataset that we created for coverage-aware retrieval. It was produced by augmenting data from from Researchy Questions~\cite{Rosset2024-hq}. The dataset contains 80K long training queries, each of which has multiple sub-questions and the LLM-judged relevance. 
    We further sample the ``pseudo'' positive and negative documents based on the coverage scores (See Section~\ref{sec:scope}). Each query has about 5 positive documents and 16 negatives.
    \item \textbf{SCOPE-flatten} is a variant derived from the SCOPE dataset. Instead of aggregating the sub-question relevance judgments for coverage, we treat each sub-question as an independent query and flatten the hierarchical structure in Researchy Questions.
    We directly use the LLM judgment rating of 5 as positive and 1 as negative, resulting in 532K training pairs for relevance ranking supervision.
\end{itemize}

\paragraph{Bi-Encoder Backbone.} 
For fair comparison, we choose ModernBERT~\cite{Warner2024-et} as the bi-encoder backbone over others because it has a longer effective input length of 8192 and supports flash-attention, compared to other neural passage retrieval models. 
We use a pre-trained checkpoint\footnote{https://huggingface.co/nomic-ai/modernbert-embed-base-unsupervised} on from Nomic-AI~\cite{nussbaum2025nomic} as initialization, avoiding the high cost of large-scale pre-training~\cite{Lee2019-nt}.
Thus, we must inherit the same setups from the pre-trained models, such as the prefix template, pooling, and similarity calculation as described in Eq.~\eqref{eq:prefix}.
We note that these settings are constraints rather than optimal design decisions for coverage-aware retrieval. 
We leave the exploration of alternative settings better suited to coverage-aware retrieval as our future work.

\paragraph{Training Configurations.} To produce encoders with a coverage-aware ranking capability on top of the relevance-based ranking, we employ two-stage finetuning as mentioned in Section~\ref{sec:rel-pft}:
We first finetune on MSMARCO for 3 epochs and further finetune with coverage-based training on SCOPE for additional 3 epochs.
All the finetuning is done with an effective query batch size of 64 with 8 documents (1 positive and 7 negatives), resulting in a total training document size of 512 (= 64 x 8), including in-batch negatives.
The learning rate is set as $10^{-4}$. Maximum query and document length are set to 180 and 512, respectively.
We share the same score temperature of 0.02 for both learning objectives $\mathcal{L}_{CovCon}$ and $\mathcal{L}_{CovDistil}$, and set the $\lambda_{CD}$ as 0.1. 

\subsection{Evaluation}
To validate the effectiveness of coverage, we use evaluation metrics for diversification and nugget coverage.
Specifically, given a ranking list, we adopt the coverage-based metrics: $\alpha$-nDCG@10 and $Cov$@10 (Subtopic Recall), which require the annotated nuggets as judgments, aiming at optimizing the context for downstream long-form RAG tasks like report generation. 
As described in Table~\ref{tab:datasets}, in this study, we evaluate CoveR on three evaluation datasets:
\begin{itemize}
    \item \textbf{NeuCLIR'24 Report Generation (ReportGen)~\cite{Dawn2025-sg}}. 
    The dataset is made for retrieval-augmented report generation. It has 19 evaluation queries and 7,049 human-annotated nugget labels. Each nugget is attached to a unique nugget question. 
    \item \textbf{CRUX Multi-Document Summarization (CRUX)~\cite{Ju2025-oz}}. It has two subsets: \textbf{DUC04} and \textbf{Multi-News}, which have 50 and 100 queries, respectively. 
    The nuggets are derived from the corresponding human-written summary.
\end{itemize}
We report both the relevance-based metrics: nDCG and Prec, as well as the coverage-based metrics: $\alpha$-nDCG and $Cov$. All metrics are computed with the cut off at 10.

Additionally, we evaluate common relevance-based retrieval tasks, BEIR~\cite{Thakur2021-cc}, which contains 13 diverse tasks across different domains. 
This evaluation provides model's performance in terms of out-of-domain relevance-based retrieval capability. Following standard practice with BEIR, the reported metric is nDCG@10.

\subsection{Baselines}
We compare CoveR with various retrieval models, including \textbf{BM25}, \textbf{Nomic-Embed}~\cite{nussbaum2025nomic},\footnote{nomic-ai/modernbert-embed-base} and \textbf{Qwen3-Embed}~\cite{qwen3embedding}\footnote{Qwen/Qwen3-Embedding-0.6B} with 0.6B and 8B.
However, these models often differ in parameter size, architecture, and training resources, so we refer to these methods as external baselines.
To this end, we add our own internal baselines for more controlled comparisons; these variants are categorized into two groups: one only finetuned with MSMARCO (\textbf{Rel.}) and one with SCOPE (\textbf{CoveR}). 
We also evaluate variants with an additional pre-finetuning stage on different relevance-based training data (\textbf{CoveR (pFT)}).
All the training is initialized with a weakly-supervised pre-trained retriever we refer to as \textbf{Unsup.} to reflect the name of the ModernBERT checkpoint used.\footnote{nomic-ai/modernbert-embed-base-unsupervised}

The other baselines include diversity ranking methods and query decomposition as our additional baselines. For example, the Maximum Marginal Relevance~\textbf{(MMR)}~\cite{Carbonell1998-gf} and the multi-query retrieval \textbf{(MultiQ)} (i.e., generate sub-queries then retrieve) using different aggregation strategies such as Reciprocal Rank Fusion \textbf{(RRF)}~\cite{cormack2009reciprocal}, Similarity Summation \textbf{(SimSum)} or Round Robin \textbf{(RRB)}.
The synthetic question generation is performed by Qwen2.5-7B-Instruct with 10 sub-questions (See Figure~\ref{fig:sub-q-gen}), and all the post-aggregations are built on top of our internal baselines fine-tuned with MSMARCO (i.e., Rel.)

\begin{table*}[]
    \centering
    \caption{
    Empirical results on nugget-based retrieval evaluation. \emph{Relevance} is measured by Precision@10 and nDCG@10, while \textit{coverage} is measured by $\alpha$-nDCG@10 and $Cov$@10. The upper block shows the external baselines and the others are our internal baselines using same backbone and training configurations.
    Statistical significance test is assessed via paired t-tests comparing against the Unsupervised$^u$ and Relevance$^r$ baselines, with results denoted by superscripts indicating $p<0.05$.
    }
    \begin{tabular}{lr cc cc cc}
\toprule
& 
& \multicolumn{2}{c}{\textbf{NeuCLIR24 ReportGen}}
& \multicolumn{2}{c}{\textbf{CRUX DUC04}}
& \multicolumn{2}{c}{\textbf{CRUX Multi-News}} \\
\cmidrule(lr){3-4} \cmidrule(lr){5-6} \cmidrule(lr){7-8} 
Model  & Size
& P / nDCG & $\alpha$-nDCG / $Cov$ 
& P / nDCG & $\alpha$-nDCG / $Cov$ 
& P / nDCG & $\alpha$-nDCG / $Cov$  \\
\midrule                                                
    \multicolumn{8}{l}{\textit{External Sparse Retrieval Baselines}} \\
\midrule    
BM25        & -     & 65.3 / 67.7 & 53.0 / 64.1 & 51.4 / 53.0 & 44.5 / 54.4 & 26.1 / 41.2 & 44.2 / 46.2 \\
SPLADE-v3   & 110M  & 81.6 / 83.1 & 62.9 / 73.7 & 68.0 / 70.4 & 55.8 / 62.4 & 34.2 / 50.7 & 51.7 / 53.6 \\
\midrule
\multicolumn{8}{l}{\textit{External Dense Retrieval Baselines}} \\
\midrule
Nomic-Embed & 149M  & 79.5 / 81.7 & 57.1 / 65.0 & 65.4 / 66.8 & 53.2 / 58.8 & 35.4 / 51.4 & 52.6 / 55.2 \\
Qwen3-Embed & 0.6B  & 81.6 / 83.6 & 58.4 / 68.5 & 69.4 / 72.0 & 55.9 / 61.4 & 35.0 / 52.4 & 56.8 / 57.0 \\
Qwen3-Embed & 8B    & 86.8 / 88.6 & 62.7 / 69.5 & 73.8 / 75.8 & 60.8 / 66.4 & 37.4 / 55.3 & 59.7 / 60.9 \\
\midrule
    \multicolumn{8}{l}{\textit{Dense Retrieval Models and Internal Baselines with ModernBERT-base (149M) }} \\
\midrule    
Unsupervised (Unsup.) & & 74.2 / 77.6 & 49.7 / 58.9 & 62.2 / 64.5 & 49.8 / 56.3 & 36.4 / 51.8 & 50.5 / 53.5 \\
Relevance (Rel.)      & & 69.5 / 72.3 & 45.8 / 55.4 & 62.8 / 64.8 & 51.9 / 58.4 & 33.9 / 48.7 & 49.5 / 51.8 \\
+ MultiQ (RRB)        & & 58.4 / 62.6 & 46.3 / 55.0 & 54.2 / 57.4 & 47.0 / 55.7 & 28.2 / 41.4 & 44.4 / 48.5 \\
+ MultiQ (RRF)        & & 63.7 / 68.1 & 50.3 / 58.2 & 58.8 / 61.3 & 48.8 / 57.1 & 30.8 / 44.0 & 45.5 / 49.6 \\
+ MultiQ (SimSum)     & & 70.5 / 73.5 & 50.2 / 55.4 & 57.6 / 58.4 & 47.1 / 56.3 & 31.0 / 42.6 & 43.4 / 49.3 \\
MMR (${\lambda=.99}$) & & 68.4 / 71.3 & 45.6 / 55.4 & 63.0 / 64.8 & 52.1 / 59.1 & 33.6 / 48.5 & 49.3 / 51.6 \\
\midrule
CoveR (w/o pFT)          & & \B{84.2} / \B{86.4} & 58.4 / 67.5 & \B{69.4}$^{ur}$ / \B{72.1}$^{ur}$ & 56.2$^{ur}$ / 61.8$^{ur}$ & 36.8$^{ur}$ / 54.3$^{ur}$ & 56.2$^{ur}$ / 57.4$^{ur}$ \\
CoveR (pFT on SCOPE-flt) & & 82.6 / 84.5 & \B{60.2}$^r$ / 67.3$^r$ & 62.2 / 65.2$^{ur}$ & 53.5$^{ur}$ / 58.2 & 36.0$^{ur}$ / 53.1$^{ur}$ & 55.2$^{ur}$ / 56.6$^{ur}$ \\
CoveR (pFT on MS)        & & 81.1 / 84.0 & 57.7$^r$ / 66.9 & 68.4$^{ur}$ / 71.0$^{ur}$ & \B{57.6}$^{ur}$ / \B{62.7}$^{ur}$ & \B{38.0}$^{r}$ / \B{55.6}$^{ur}$ & \B{58.4}$^{ur}$ / \B{59.0}$^{ur}$ \\
\bottomrule
    \end{tabular}
    \label{tab:coverage-base-eval}
\end{table*}

\section{Experimental Results}
In this section, we report results on both the relevance-based ranking and nugget-based evaluation datasets. We empirically assess the two retrieval capabilities of CoveR, along with analyses on the importance of SCOPE training data.

\subsection{Effectiveness on Nugget-based Benchmarks}
In Table~\ref{tab:coverage-base-eval}, we present nugget-based retrieval performance across three benchmarks: NeuCLIR24 ReportGen, CRUX DUC04 and CRUX Multi-News, evaluated via two complementary dimensions: relevance (P@10 and nDCG@10) and nugget coverage ($\alpha$-nDCG@10 and $Cov$@10).

Among all datasets, external baselines such as SPLADE-v3, Nomic-Embed, and Qwen3-Embed achieve strong relevance performance. 
Qwen3-Embed-8B consistently achieves the highest performance on relevance-based metrics in all the benchmarks.
However, for the coverage-based metrics, we observe that SPLADE-v3 achieves higher coverage-based scores than the larger retrieval models (62.9/73.7 vs. 62.7/69.5 on NeuCLIR24 ReportGen), even though Qwen3-Embed-8B has significantly better relevance-based ranking effectiveness.
As for the comparisons of the other two models, they show similar results, with Qwen3-Embed-0.6B being slightly better, likely because of the larger model size.

In the lower block, we report internal baselines that shares the same ModernBERT backbone that CoveR uses. 
First, we observe that the original pre-trained checkpoint (Unsup.\footnote{While Nomic named this checkpoint unsupervised, note that it is trained with a large amount of weakly-supervised query-document pairs.}) exhibits decent retrieval capabilities, and sometimes even outperforms Nomic-Embed, which was finetuned with a larger amount of supervised datasets (e.g., nDCG 51.8 vs. 51.4 on CRUX Multi-News). This indicates that relevance-based supervision is beneficial, but the improvement may be small.
Similar to this observation, we found that our relevance-based trained model variant Rel. (i.e., Unsupervised + MSMARCO finetuning) has noticeable drops on NeuCLIR ReportGen and CRUX Multi-News, suggesting that the relevance-only supervision is insufficient for nugget-based evaluation benchmarks and drives the model to focus on a narrow view of relevance. We hypothesize that this is due to the misaligned relevance definition in the nugget-based evaluation. 
We also observe that all the heuristic diversification approaches hurt the effectiveness; they are mostly inferior to the original base model (Rel.), except for a small improvement in terms of $\alpha$-nDCG on NeuCLIR24 ReportGen (e.g., MultiQ (*) 46.3/50.3/50.2 vs. 45.8).

In contrast, CoveR outperforms all the internal baselines by implicitly modeling coverage signals during retrieval.
Particularly, compared to Unsup. and Rel., CoveR consistently outperforms these baselines across all evaluation datasets and metrics, regardless of whether pre-finetuning is used.
Moreover, CoveR even performs on par with Nomic-Embed and Qwen3-0.6B, which are trained on larger scale of data and with larger model backbones.

However, the impact of pre-finetuning is mixed. Considering relevance metrics, omitting pre-finetuning leads to slightly higher effectiveness on NeuCLIR ReportGen and CRUX DUC04 while being slightly lower on CRUX Multi-News.
Considering nugget coverage metrics, CoveR with MSMARCO pre-finetuned version is better in 4/6 cases, with NeuCLIR24 ReportGen being the exception.
These results indicate that pre-finetuning before coverage training is not essential for coverage, but it can help improve effectiveness for some use cases.

Finally, across all the different evaluation measurements, we see a distinction between the two families of metrics. We observe that the variance of coverage-based metrics across all systems is generally smaller than that of relevance-based metrics. For example, for NeuCLIR24 ReportGen, the variance of metrics Precision, nDCG, $\alpha$-nDCG, and $Cov$ are 0.8, 0.7, 0.4, and 0.4\%.
This gap points out that the relevance-based ranking is still important to the coverage-based ranking.
However, the increase in relevance ranking capability might not be fully correlated to the gains in nugget-based evaluation coverage when comparing competitive model candidates. This requires more exploration along with learning the interactions between the two capabilities to satisfy the upcoming retrieval demands.

\begin{table*}[]
    \centering
    \caption{
    Empirical results on relevance-based retrieval evaluation using BEIR datasets. 
    We report nDCG@10 on 13 of them and the average is reported at the first column.
    The external and internal baselines are in the upper and lower parts, respectively. 
    The values with bold font indicate the best performance within each group. 
    }
    \begin{tabular}{lc|ccccccccccccc}
\toprule
    Model & Avg. & arg	& cli  & scif  & tre	 & web	& dbp  & fev  & fiq  & hot  & nfc  & nq   & quo  & scid  \\
\midrule
    BM25                   & 42.2 & 39.7 & 16.5 & 67.9 & 39.5 & 44.2 & 31.8 & 65.1 & 23.6 & 63.3 & 32.2 & 30.5 & 78.9 & 14.9 \\
    SPLADE-v3              & 51.7 & 50.9 & 23.3 & 71.0 & 74.8 & 29.3 & 45.0 & 79.6 & 37.4 & 69.2 & 35.7 & 58.6 & 81.4 & 15.8 \\
    Nomic-Embed            & 52.0 & 35.7 & 34.4 & 68.8 & 81.2 & 35.3 & 39.8 & 85.3 & 40.6 & 65.9 & 32.4 & 52.1 & 87.5 & 17.0 \\
    Qwen3-Embed            & \B{54.0} & 45.5 & 36.2 & 69.0 & 87.6 & 27.6 & 39.5 & 85.9 & 46.2 & 65.2 & 35.7 & 52.9 & 87.4 & 22.9 \\
\midrule
    Unsupervised           & 47.5     & \B{37.6} & 23.0     & \B{72.4} & 67.9     & 18.6     & 38.1     & 67.4     & \B{42.4} & 60.0 & \B{35.4} & 46.3     & \B{88.8} & 19.8 \\
    Relevance              & 50.1     & 34.0     & 26.3     & 70.3     & \B{82.0} & \B{26.5} & 39.0     & 77.3     & 39.6     & \B{63.0} & 33.5 & 55.3     & 86.2 & 18.2 \\
    CoveR (w/o pFT)        & 49.0     & 36.4     & 24.6     & 71.4     & 71.0     & 20.7     & \B{39.1} & 76.3     & 40.9     & 61.1 & 34.9     & 52.5     & 87.4 & \B{20.3} \\
    CoveR (pFT: SCOPE-flt) & 47.9     & 34.6     & \B{26.5} & 69.9     & 66.3     & 20.4     & 37.1     & \B{79.9} & 38.2     & 61.6 & 33.0     & 49.9     & 85.9 & 19.7 \\
    CoveR (pFT: MS)        & \B{50.2} & 36.0     & 26.3     & 70.9     & 78.4     & 26.3     & 39.0     & 78.9     & 39.8     & 62.2 & 34.5     & \B{55.8} & 84.9 & 19.6 \\
\bottomrule
    \end{tabular}
    \label{tab:relevance-base-eval}
\end{table*}

\subsection{Effectiveness on  Relevance Ranking}
Aside from the emerging coverage-aware retrieval capability of CoveR, we evaluate the relevance ranking performance using BEIR datasets.
Table~\ref{tab:relevance-base-eval} compares our proposed method with the same external and internal baselines.
We observe that relevance-based finetuning with MSMARCO can boost the overall average nDCG@10 of BEIR by 2.6 points, which differs from the trend that appeared in nugget-based evaluation benchmarks.
In addition, we found that CoveR pre-finetuned with MSMARCO can preserve the relevance ranking capability and even slightly outperform CoveR without pre-finetuning on 6 out of 13 datasets. 
However, CoveR pre-finetuned with SCOPE-flatten yields limited gains (47.5 to 47.9), indicating that the LLM judgment labels from SCOPE-flatten are still not as useful as MSMARCO in terms of relevance-based ranking.
We also observe the improvement when finetuning only with coverage signals (+1.5), indicating the SCOPE dataset is useful for both relevance and coverage.
In Figure~\ref{fig:two-metrics-cov}, we plot the performance of the compared retrieval systems across two different capabilities. This illustrates that coverage-based training can boost coverage scores by 5 points while maintaining a similar level of relevance ranking capability.
Finally, we found that SPLADE-v3 naturally obtains superior performance on nugget coverage. We hypothesize that this is due to the relevance estimation process of sparse retrieval, which incurs the hierarchical information of nuggets in the token expansion. We consider the coverage-aware sparse retrieval as our future work.

\begin{figure}[t]
    \centering
    \includegraphics[width=.95\columnwidth]{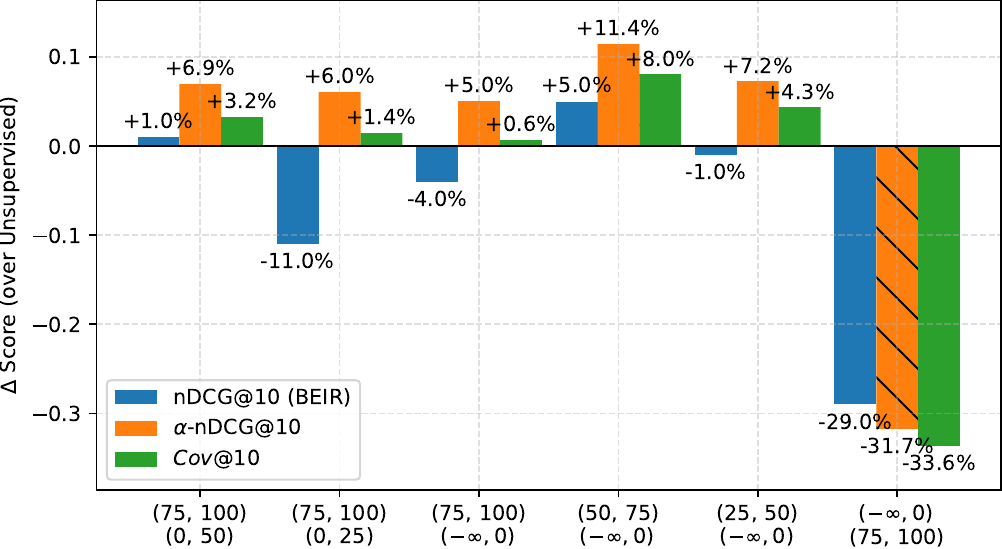}
    \caption{
    The performance changes over \textit{Unsupervised} baseline with different sampling ranges for positive and negatives. The criteria of two ranges are based on the estimated coverage score of documents. The upper parenthesis ($\alpha_1, \alpha_2$) is the range for high coverage documents; the lower parenthesis ($\beta_1, \beta_2$.) is the range for low coverage documents.
    }
    \label{fig:cov-sampling}
\end{figure}

\begin{table*}[h]
\caption{The reference bounds for three nugget-based retrieval evaluation using the provided sub-questions (oracle) in the retrieval systems.}
    \label{tab:oracle}
    \centering
    \begin{tabular}{l cc cc cc}
\toprule
& \multicolumn{2}{c}{\textbf{NeuCLIR24 ReportGen}}
& \multicolumn{2}{c}{\textbf{CRUX DUC04}}  
& \multicolumn{2}{c}{\textbf{CRUX Multi-News}}  \\
\cmidrule(lr){2-3} \cmidrule(lr){4-5} \cmidrule(lr){6-7} 
Retrieve-then-aggregate  
& P / nDCG & $\alpha$-nDCG / $Cov$ 
& P / nDCG & $\alpha$-nDCG / $Cov$ 
& P / nDCG & $\alpha$-nDCG / $Cov$  \\
\midrule
Relevance           & 69.5 / 72.3 & 45.8 / 55.4 & 62.8 / 64.8 & 51.9 / 58.4 & 33.9 / 48.7 & 49.5 / 51.8 \\
+ MultiQ (SimSum)   & 70.5 / 73.5 & 50.2 / 55.4 & 57.6 / 58.4 & 47.1 / 56.3 & 31.0 / 42.6 & 43.4 / 49.3 \\
+ OracleQ (SimSum)  & 75.3 / 77.4 & \B{63.0} / \B{72.0} & 75.2 / 76.3 & 61.8 / 69.4 & \B{42.1} / 61.7 & 67.3 / 65.6 \\
+ MultiQ (RRF)      & 63.7 / 68.1 & 50.3 / 58.2 & 58.8 / 61.3 & 48.8 / 57.1 & 30.8 / 44.0 & 45.5 / 49.6 \\
+ OracleQ (RRF)     & 65.8 / 68.9 & 56.4 / 65.6 & \B{82.4} / \B{84.5} & \B{69.8} / \B{76.6} & 41.3 / 63.2 & \B{69.4} / \B{67.9} \\
CoveR (pFT on MS)   & \B{81.1} / \B{84.0} & 57.7 / 66.9 & 68.4 / 71.0 & 57.6 / 62.7 & 38.0 / 55.6 & 58.4 / 59.0 \\
\bottomrule
    \end{tabular}
\end{table*}

\begin{table}[t]
    \caption{Evaluation results of CoveR variants across different weight of coverage distillation ($\lambda_{CD}$).}
    \label{tab:ablation}
    \centering
    \label{tab:variants}
    \begin{tabular}{ll cc}
    \toprule
                   & &  \textbf{BEIR}              & \textbf{CRUX DUC04}\\
    Query type     & $\lambda_{CD}$ & nDCG & P / nDCG / $\alpha$-nDCG / $Cov$ \\ 
    \midrule
    Re-Constructed & 0.0  & 50.1  & 68.4 / 71.1 / 57.2 / 61.3 \\
                   & 0.1  & 50.2  & 68.4 / 71.0 / 57.6 / 62.7 \\
                   & 0.25 & 48.3  & 68.0 / 70.7 / 57.8 / 61.4 \\ 
    Original       & 0.0  & 49.3  & 66.6 / 68.6 / 55.5 / 61.2 \\
                   & 0.1  & 50.0  & 66.0 / 68.4 / 55.4 / 60.7 \\
    \bottomrule
    \end{tabular}
    
\end{table}

\subsection{Empirical Analysis}
To better understand our proposed design choices, we analyze the CoveR through answering the following research questions.

\subsubsection{How do sampled positive/negative documents affect CoveR’s relevance- and coverage-based ranking capability?}\label{sec:ana-cov-sample}
To examine how coverage-aware sampling affects ranking capability, we compare multiple configurations for constructing positive and negative training samples under different sampling ranges, as described in Eq.~\eqref{eq:cov-sampling}.
Figure~\ref{fig:cov-sampling} shows the improvements (or deterioration) over the initialized performance (Unsup.). 
All the compared configurations include two ranges controlled by $\alpha$ and $\beta$.
We evaluate the retrieval performance using average nDCG@10 on BEIR and average coverage-based metrics on three nugget-based datasets.

Empirically, expanding the positive sampling range leads to consistent gains in coverage-based metrics (the marked bars). Sampling positive from documents that have greater than 50\% of coverage yields the optimal gains. 
We found the range (75, 100) has weaker gains because these documents are rare, resulting in inadequate positive examples and therefore collapsing the relevance capability.
As for the negatives, we found that mixing low-coverage documents with zero-coverage documents can enhance the learning, which serves as a harder negative. 
To further validate the coverage signals, we also intentionally conduct the ``reversed'' configurations by sampling positive from ($-\infty, 0)$ and negatives from (75, 100). As expected, this setting severely breaks the coverage and relevance retrieval capability.

\subsubsection{What are the reference bounds for nugget-based retrieval benchmarks?}\label{sec:ana:oracle}
To better understand the strength of coverage-aware retrieval, we compare it with an oracle multi-query retrieval setting, OracleQ, where the ``golden'' annotated nugget sub-questions are available.
In Table~\ref{tab:oracle}, we report the results on three nugget-based evaluation datasets. 
Similar to MultiQ in Table~\ref{tab:coverage-base-eval}, we directly treat the annotated sub-questions as the query and perform relevance based ranking, resulting in multiple independent ranking.
Then, we aggregate the results into a final ranking via RRF or summing all the similarity scores. 
We observe that swapping with nugget sub-questions can bring noticeable gains in terms of coverage metrics, increasing $\alpha$-nDCG by more than 10 points on NeuCLIR (with SimSum aggregation), CRUX-DUC and Multi-News (using RRF aggregation), showing that there is still ample room to improve.

This also points out a fundamental limitation of static retrieval like CoveR. Although CoveR implicitly considers coverage, it still relies on similarity estimation that ranks documents independently rather than retrieving them as a whole~\cite{Lee2025-xm, ju2026lancer}, or selecting them iteratively~\cite{trivedi2023interleaving}.
As a result, the model lacks an explicit mechanism to avoid redundancy and ensure complementary coverage across retrieved documents. 

\begin{figure}[t]
\begin{tcolorbox}[title=LLM Sub-Question Generation Prompt, myprompt]
    Instruction: Write 10 diverse sub-questions that can reveal the information required to answer the given report request. Each sub-question should be self-contained and include the necessary context. Collectively, the sub-questions should fully cover the scope of the report request. Write each sub-question within '<q>' and '</q>' tags.  
    \\ \\
    Report request: \{query\} \\
    Sub-questions: <q>
\end{tcolorbox}
\caption{The prompt used for generating sub-questions. The decomposed sub-questions are used to issue multiple searches (i.e., MultiQ).}
\label{fig:sub-q-gen}
\end{figure}

\subsubsection{What is the impact of training query types on coverage-aware retrieval?}\label{sec:ablation}
We aim to investigate how the query affects the quality of coverage-based training.
We compare the reconstructed query to the original query in Researchy Questions (see Table~\ref{tab:scope-example}) and report the evaluation results on BERT and CRUX DUC04.
Table~\ref{tab:ablation} shows the models trained with re-constructed query demonstrate the better effectiveness on DUC04 while the relevance-based ranking effectiveness remain similar. We hypothesize that the re-constructed query can guide the retrieval model with lexical matching, so that retrieval models can capture more signals for coverage instead of solely relying on semantic similarity.
 
\subsubsection{What is the optimal weight for coverage distillation?}
To determine a more suitable weight for mixing two learning objectives, we vary the $\lambda_{CD}$ within the range [0, 0.1, 0.25]. In Table~\ref{tab:ablation}, we found the weight $0.1$ achieves the optimal across different metrics. However, when using original query, the trend is inverted on CRUX DUC04; in this setting the coverage contrastive learning alone ($\lambda_{CD}=0.0$) is sufficient to achieve improved coverage-based metrics.

\section{Conclusion}
In this work, we presented CoveR, a coverage-aware neural retriever designed for long-form RAG scenarios. 
By learning from coverage-based signals derived from sub-questions (i.e., Coverage contrastive and Coverage distillation), CoveR reshapes the representation space to better reflect which documents collectively satisfy multiple information needs. 
To support such learning methods, we introduce SCOPE dataset, which contains coverage signals augmented via LLMs. 
Our empirical evaluation shows that CoveR achieves better performance in terms of coverage while preserving the original relevance ranking capability by retrieval pre-training. 
This is particularly important for the future development of retrieval for RAG. 
Some promising future directions include integrating CoveR as one of a searching strategies into agentic pipeline to tackle information needs with different requirement of nuggets.

\begin{acks}
This research was supported by the \href{https://hybrid-intelligence-centre.nl}{Hybrid Intelligence Center}, a 10-year program funded by the Dutch Ministry of Education, Culture and Science through the Netherlands Organisation for Scientific Research, project VI.Vidi.223.166 of the NWO Talent Programme which is (partly) financed by the Dutch Research Council (NWO) and NWO project NWA.1389.20.\-183.
We also acknowledge the Dutch Research Council for awarding this project access to the LUMI supercomputer, owned by the EuroHPC Joint Undertaking, hosted by CSC (Finland) and the LUMI consortium through project NWO-2025.040.
    
The authors acknowledge the peoples of the Woi Wurrung and Boon Wurrung language groups of the eastern Kulin Nation on whose unceded lands ACM SIGIR 2026 was hosted. We pay our respects to their Elders past and present, and extend that respect to all Aboriginal and Torres Strait Islander peoples today and their continuing connection to land, sea, sky, and community.    
\end{acks}


\bibliographystyle{ACM-Reference-Format}
\balance
\bibliography{custom}

\appendix

\end{document}